\documentclass[preprint,showpacs,showkeys,preprintnumbers,amsmath,amssymb,superscriptaddress]{revtex4}
\usepackage{graphicx}
\usepackage{dcolumn}
\usepackage{bm}

\begin{document}

\title{Entropic gravity versus gravitational pseudotensors in static spherically symmetric spacetimes}

\author{S. Hamid Mehdipour}

\email{mehdipour@liau.ac.ir}

\affiliation{Department of Physics, College of Basic Sciences,
Lahijan Branch, Islamic Azad University, P. O. Box 1616, Lahijan,
Iran}

\date{\today}

\begin{abstract}
We present some well-known energy-momentum complexes and evaluate
the gravitational energy associated with static spherically
symmetric spacetimes. In fact, the energy distribution of the
aforementioned gravitational background that is contained in a
two-sphere of radius $r$ shows that a test particle situated at a
finite distance $r$ experiences the gravitational field of the
effective gravitational mass. In addition, we apply Verlinde's
entropic gravity to find the emergent gravitational energy on static
spherically symmetric screens. In this setup, we find that the
energy distribution in the prescription of M{\o}ller is similar to
the energy derived from the emergent gravity, while other
prescriptions give the different results. This result may confirm
the argument of Lessner who argues that M{\o}ller's definition of
energy is a powerful concept of energy in General Relativity.
\end{abstract}


\maketitle

\section{\label{sec:1}Introduction}
The notion of energy has been an issue of extensive research since
the beginning of Einstein's theory of General Relativity (GR).
Einstein \cite{Ein} was the first to consider a locally conserved
formula for energy-momentum complexes including the contribution
from the gravitational field energy. He obtained an expression for
the energy-momentum complex by introducing the gravitational field
pseudotensor. This idea followed by a surge of interest and the
various prescriptions for the energy-momentum complexes as tools to
determine energy-momentum distributions were suggested
\cite{Tol,Pap,Lan,Ber,Gol,Wei}. These prescriptions were confined to
evaluate the energy and momentum in quasi-Cartesian coordinates.
Afterwards, a new description for the energy-momentum complex was
proposed by M{\o}ller \cite{Mol} which is not restricted to
quasi-Cartesian coordinates and furthermore it provides a powerful
concept of energy and momentum in GR \cite{les}. Nevertheless, the
topic of energy-momentum complexes has been fundamentally disputed
in the context of GR (see \cite{vage} and the references therein).
There are many unanswered questions regarding the energy and
momentum localization in the literature (see for instance
\cite{Xul}). There have been a lot of attempts to obtain a
well-behaved expression for local or quasi-local energy-momentum and
a number of studies have been performed on this debatable issue
\cite{Vir}.

In this paper, we will evaluate the gravitational energy associated
with static spherically symmetric (SSS) spacetimes. We employ two
approaches. The first one is the pseudotensor approach and the other
one is the entropic approach. It is evident that there exists a link
between gravity and thermodynamics. Jacobson \cite{jac} has
exhibited that the Einstein field equations of GR are derived from
the first law of thermodynamics. Padmanabhan \cite{pad} has applied
the equipartition law of energy and the holographic principle to
present a thermodynamic explanation of gravity (see also
\cite{pad2}). A recent work by Verlinde \cite{ver} has suggested a
novel approach to interpret the gravity as an entropic force owing
to alterations in the information connected to the positions of
material bodies. In this approach, one could imagine an emergent
phenomenon for the origin of Newtonian gravity. This theory
indicates that the gravitational interaction arises from the
statistical behavior of microscopic degrees of freedom encoded on a
holographic screen. The entropic approach has recently given an
impetus for more study in the literature \cite{ent}.

The paper is organized as follows. In Sec. \ref{sec:2}, using the
energy-momentum definitions of Einstein, Landau-Lifshitz, Weinberg,
Papapetrou and M{\o}ller, we calculate the energy distribution of
the SSS spacetimes in a generic form, respectively. In Sec.
\ref{sec:3}, we study Verlinde's idea about the temperature and the
energy on the holographic screens for generic spherically symmetric
surfaces. Finally, we present a summary in Sec. \ref{sec:4}. In this
work, we will choose to write spacetime indices using the Greek
alphabet, and space indices using the Latin alphabet. We will also
choose to use natural units, where $\hbar = c = G = k_B = 1$.

\section{\label{sec:2}Energy-Momentum Complexes}
The energy-momentum conservation in the context of GR can be written
as
\begin{equation}
\label{mat:1} \nabla _{\mu}T^ {\mu}_ {\nu} = 0,
\end{equation}
where $T^{\mu}_{\nu}$ is the symmetric energy-momentum tensor
containing the matter and all non-gravitational fields. Einstein
\cite{Ein} conjectured a energy-momentum complex
${\cal{T}}^{\mu}_{\nu}$ containing the matter, all non-gravitational
fields and the gravitational field such that it obeys a conservation
law in the form of a divergence in the following form
\begin{equation}
\label{mat:2}{\cal{T}}^{\mu}_{\nu \,\, , \mu}=0,
\end{equation}
with
\begin{equation}
\label{mat:3}{\cal{T}}^{\mu}_{\nu}= \sqrt{-g} (T_{\nu}^{\mu}
+t_{\nu}^{\mu}),
\end{equation}
where comma indicates partial differentiation and $g$ is the
determinant of the metric tensor $g_{\mu\nu}$. The expression
$t_{\nu}^{\mu}$ exhibits the energy-momentum pseudotensor which is a
nontensorial quantity to describe the gravitational field energy. We
can also write the energy-momentum complex in the following form
\begin{equation}
\label{mat:4}{\cal{T}}^{\mu}_{\nu}=\theta^{\mu\lambda}_{\nu \,\,\,\,
, \lambda},
\end{equation}
where $\theta^{\mu\lambda}_{\nu}$ is denoted as superpotential
components which are functions of the metric tensor and its first
order derivatives. It is obvious that the energy-momentum complex
does not have a unique definition due to the fact that one may
always add a quantity with a zero divergence to the expression
${\cal{T}}^{\mu}_{\nu}$.

We are interested in computing the energy distribution associated
with the SSS gravitational background, which is contained in a
two-sphere of radius $r$. Thus, the background metric is supposed to
be a generic SSS solution as follows:
\begin{equation}
\label{mat:10}ds^2=-A(r)dt^2+ B(r)dr^2+r^2 d\Omega^2,
\end{equation}
where $A$ and $B$ are arbitrary functions of the radial coordinate
and $d\Omega^2=d\theta^2+\sin^2\theta d\varphi^2$ gives the standard
line element on the unit two-sphere. For carrying out the
calculations with Einstein, Landau-Lifshitz, Weinberg and Papapetrou
energy-momentum complexes, we require to reexpress the SSS metric in
quasi-Cartesian coordinates. Transforming (\ref{mat:10}) to
Cartesian terms according to $x= r \sin\theta\cos\varphi$, $y= r
\sin\theta \sin\varphi$, and $z=r\cos\theta$, one gets the metric
\begin{equation}
\label{mat:11}ds^2=-A(r)dt^2+dx^2+dy^2+dz^2+\frac{B(r)-1}{r^2}\left(xdx+ydy+zdz\right)^2,
\end{equation}
where $r^2=x^2+y^2+z^2$. In the following subsections, we will
present some well-known energy-momentum complexes for obtaining the
gravitational energy in SSS spacetimes.

\subsection{\label{sec:2.1}Energy distribution in Einstein's
prescription} Einstein's energy-momentum complex \cite{Ein} has the
form
\begin{equation}
\label{mat:5}\theta^\mu_\nu = \frac{1}{16\pi}h^{\mu\lambda}_{\nu~\,
,\,\lambda},
\end{equation}
where Einstein's superpotential $h^{\mu\lambda}_{\nu}$ is given by
\begin{equation}
\label{mat:6}h^{\mu\lambda}_{\nu}
=\frac{1}{\sqrt{-g}}g_{\nu\sigma}\left[-g\left(g^{\mu\sigma}g^{\lambda\kappa}-g^{\lambda\sigma}g^{\mu\kappa}
\right)\right]_{,\,\kappa},
\end{equation}
with the antisymmetric property
\begin{equation}
\label{mat:7}h^{\mu\lambda}_{\nu}=-h^{\lambda\mu}_{\nu}.
\end{equation}
The energy in Einstein's prescription for a four-dimensional
background is given by
\begin{equation}
\label{mat:8} E=\int\int\int \theta_0^0 dx^1dx^2dx^3,
\end{equation}
where $\theta_0^0$ is the energy density of the total physical
system including gravitation. The integrals in Eq. (\ref{mat:8}) are
extended over all space for $x^0 = const$. Using Gauss's theorem,
the energy component is equal to
\begin{equation}
\label{mat:9} E=\frac{1}{16\pi}\int\int h^{0\, i}_0 \, n_i \, dS,
\end{equation}
where $n_i=x_i/r$ is the outward unit normal vector over an
infinitesimal surface element $dS$. Using Eq. (\ref{mat:9}) and
evaluating the integrals over the surface of two-sphere of radius
$r$, the energy distribution associated with the generic SSS metric
in Einstein's formulation is found to be
\begin{equation}
\label{mat:12}E_E=\frac{rA(B-1)}{2\sqrt{AB}}.
\end{equation}

\subsection{\label{sec:2.2}Energy distribution in Landau-Lifshitz's prescription}
The energy and momentum in the prescription of
Landau-Lifshitz \cite{Lan} is given by
\begin{equation}
\label{mat:13} L^{\mu\nu} = \frac{1}{16\pi}{\cal
S}^{\mu\nu\lambda\kappa} _ {\,~~\quad ,\lambda\kappa},
\end{equation}
with
\begin{equation}
\label{mat:14}{\cal S}^{\mu\nu\lambda\kappa} =
-g\left(g^{\mu\nu}g^{\lambda\kappa} -
g^{\mu\lambda}g^{\nu\kappa}\right),
\end{equation}
where $L^{\mu\nu}$ is symmetric with respect to its indices.
Landau-Lifshitz's superpotential ${\cal S}^{\mu\nu\lambda\kappa}$
has symmetries similar to the curvature tensor. The energy in the
Landau-Lifshitz prescription for a four-dimensional background is
given by
\begin{equation}
\label{mat:15} E=\int\int\int L^{00}dx^1dx^2dx^3,
\end{equation}
where $L^{00}$ is the energy density component. Using Gauss's
theorem, the energy component is
\begin{equation}
\label{mat:16} E=\frac{1}{16\pi}\int\int {\cal
S}^{00i\kappa}_{~\quad ,\kappa} \, n_i \, dS.
\end{equation}
Using the metric (\ref{mat:11}), we get the energy distribution in
Landau-Lifshitz's definition in the following form
\begin{equation}
\label{mat:17}E_{LL}=\frac{r}{2}(B-1).
\end{equation}

\subsection{\label{sec:2.3}Energy distribution in Weinberg's prescription}
The Weinberg's energy-momentum complex \cite{Wei} is expressed as
\begin{equation}
\label{mat:18}W^{\mu\nu} =\frac{1}{16\pi}\Delta^{\mu\nu\lambda}_
{~\quad, \lambda},
\end{equation}
where Weinberg's superpotential $\Delta^{\mu\nu\lambda}$ is
antisymmetric on its first pair of indices which defines as
\begin{equation}
\label{mat:19}\Delta^{\mu\nu\lambda} = \partial^\mu h^\kappa_\kappa
\eta^{\nu\lambda} - \partial^\nu h^\kappa_\kappa \eta^{\mu\lambda} -
\partial_\kappa h^{\kappa\mu} \eta^{\nu\lambda}
+ \partial_\kappa h^{\kappa\nu} \eta^{\mu\lambda} + \partial^\nu
h^{\mu\lambda} - \partial^\mu h^{\nu\lambda},
\end{equation}
where $\partial_\mu\equiv\partial/\partial x^\mu$,
$\partial^\mu\equiv\partial/\partial x_\mu$ and $h_{\mu\nu}$ shows
the symmetric tensor defined as
$h_{\mu\nu}=g_{\mu\nu}-\eta_{\mu\nu}$, where $\eta_{\mu\nu}$ is the
Minkowski metric. The energy in Weinberg's prescription for a
four-dimensional background is given by
\begin{equation}
\label{mat:20} E=\int\int\int W^{00} dx^1dx^2dx^3,
\end{equation}
where $W^{00}$ is the energy density component. Using Gauss's
theorem, one has
\begin{equation}
\label{mat:21} E=\frac{1}{16\pi}\int\int \Delta^{i00} \, n_i \, dS.
\end{equation}
The energy distribution connected to the generic SSS metric in
Weinberg's formulation becomes the same as the energy derived from
Landau-Lifshitz's prescription, i.e.
\begin{equation}
\label{mat:22}E_W=E_{LL}=\frac{r}{2}(B-1).
\end{equation}

\subsection{\label{sec:2.4}Energy distribution in Papapetrou's prescription}
The energy and momentum in the prescription of Papapetrou \cite{Pap}
takes the form
\begin{equation}
\label{mat:23}\Omega^{\mu\nu} = \frac{1}{16\pi}
N^{\mu\nu\lambda\kappa} _{\,~~\quad,\lambda\kappa},
\end{equation}
with
\begin{equation}
\label{mat:24}N^{\mu\nu\lambda\kappa} =
\sqrt{-g}\left(g^{\mu\nu}\eta^{\lambda\kappa} -
g^{\mu\lambda}\eta^{\nu\kappa} + g^{\lambda\kappa}\eta^{\mu\nu} -
g^{\nu\kappa}\eta^{\mu\lambda}\right),
\end{equation}
where $ N^{\mu\nu\lambda\kappa}$ is Papapetrou's superpotential and
is symmetric on its first pair of indices. The energy in the
Papapetrou prescription for a four-dimensional background is given
by
\begin{equation}
\label{mat:25} E=\int\int\int \Omega^{00}dx^1dx^2dx^3,
\end{equation}
where $\Omega^{00}$ represents the energy density component. Using
Gauss's theorem, the energy component is
\begin{equation}
\label{mat:26} E=\frac{1}{16\pi}\int\int N^{00ij}_{~\quad ,j} \, n_i
\, dS.
\end{equation}
Using the metric (\ref{mat:11}), we get the energy component of
Papapetrou's definition in the following form
\begin{equation}
\label{mat:27}E_{P}=\frac{r}{8(AB)^{\frac{3}{2}}}\left[4A^2B(B-1)+r\left(A'B^2-ABB'-AA'B+A^2B'\right)\right],
\end{equation}
where the prime abbreviates $\partial/\partial r$.

\subsection{\label{sec:2.5}Energy distribution in M{\o}ller's prescription}
The energy-momentum complex of M{\o}ller \cite{Mol} is given by
\begin{equation}
\label{mat:28}M^{\mu}_\nu=\frac{1}{8\pi}\chi^{\mu\lambda}_{\nu
\,\,\,\, ,\lambda },
\end{equation}
where M{\o}ller's superpotential $ \chi^{\mu\lambda}_\nu $ has the
form
\begin{equation}
\label{mat:29}\chi^{\mu \lambda}_\nu=-\chi^{ \lambda\mu}_\nu=\sqrt{-
g}\left(g_{\nu \sigma, \kappa}- g_{\nu\kappa,
\sigma}\right)g^{\mu\kappa} g^{\lambda\sigma}.
\end{equation}
The energy component in M{\o}ller's prescription is given by
\begin{equation}
\label{mat:30} E=\int\int\int M_0^0 dx^1dx^2dx^3,
\end{equation}
where $M_0^0$ is the energy density component. Using Gauss's
theorem, the energy component is equal to
\begin{equation}
\label{mat:31} E=\frac{1}{8\pi}\int\int \chi^{0\, i}_0 \, n_i \, dS.
\end{equation}
Note that the calculations are not anymore confined to
quasi-Cartesian coordinates. Hence, we utilize the metric
(\ref{mat:10}) to get the energy distribution in M{\o}ller's
definition as follows:
\begin{equation}
\label{mat:32}E_{M}=\frac{r^2A'}{2\sqrt{AB}}.
\end{equation}
It would be worthwhile to denote that, the energy given by
Eqs.~(\ref{mat:12}), (\ref{mat:17}), (\ref{mat:22}), (\ref{mat:27})
and (\ref{mat:32}) is also called the effective gravitational mass
$M_{eff}$ of the spacetime under consideration. For instance, the
energy distribution in M{\o}ller's prescription given by
Eq.~(\ref{mat:32}) is in fact the energy (effective mass) of the
gravitational field that a test particle present at a finite
distance $r$ in this field experiences. Studying on the problem of
finding the effective gravitational mass was first considered by
Cohen and Gautreau \cite{coh}. Afterwards, much attention has been
devoted to this issue for different spacetimes \cite{vag}.

In the following, we will present Verlinde's entropic scenario
\cite{ver} to investigate the emergent gravitational energy on SSS
screens.

\section{\label{sec:3}Entropic gravity}
\begin{figure}[htp]
\begin{center}
\includegraphics{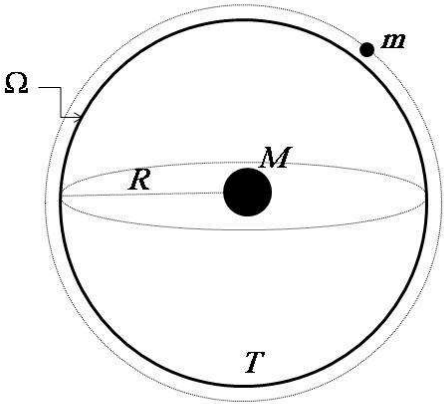}
\end{center}
\vspace{6.7 cm} \caption{\scriptsize {A test particle with mass $m$
approaches the spherical holographic screen $\Omega$. The screen
$\Omega$ possesses an equilibrium temperature $T$ and the total
equipartition energy $E$ which is located at a distance of $R$ from
the source mass $M$ at the origin. It is supposed that the energy
associated with the source mass is evenly dispersed on the screen.}}
\label{fig:1}
\end{figure}

In order to obtain the energy on a holographic screen for a generic
SSS spacetime, we should find the timelike Killing vector of the
metric (\ref{mat:10}). Using the Killing equation
\begin{equation}
\label{mat:33}\partial_\mu\xi_\nu+\partial_\nu\xi_\mu-2\Gamma^\lambda_{\mu\nu}\xi_\lambda=0,
\end{equation}
with the condition of SSS, i.e.
$\partial_0\xi_\mu=\partial_3\xi_\mu=0$, and also the infinity
condition $\xi_\mu\xi^\mu=-1$, the timelike Killing vector is
written as
\begin{equation}
\label{mat:34}\xi_\mu=\left(-A,\,0,\,0,\,0\right).
\end{equation}
To define a foliation of space, and distinguishing the holographic
screens $\Omega$ at surfaces of constant redshift, we write the
generalized Newtonian potential $\phi$ in the general relativistic
framework
\begin{equation}
\label{mat:35}\phi=\frac{1}{2}\log\left(-g^{\mu\nu}\xi_\mu\xi_\nu\right)=
\frac{1}{2}\log A,
\end{equation}
where $e^\phi$ is the redshift factor and is equal to one at the
reference point with $\phi = 0$ at infinity. Thus, the acceleration
$a^\mu$ for a particle that is placed close to the screen yields the
following:
\begin{equation}
\label{mat:36}a^\mu=-g^{\mu\nu}\nabla_\nu\phi=\left(0,\,\frac{A'}{2AB},\,0,\,0\right).
\end{equation}
The temperature on the holographic screen is given by Unruh-Verlinde
temperature that is connected to the proper acceleration of a
particle near the screen and can be written as \cite{ver}
\begin{equation}
\label{mat:37}T=-\frac{1}{2\pi}e^\phi n^\mu
a_\mu=\frac{e^\phi}{2\pi}\sqrt{g^{\mu\nu}\nabla_\mu\phi\nabla_\nu\phi},
\end{equation}
where
$n^\mu=\nabla^\mu\phi/\sqrt{g^{\mu\nu}\nabla_\mu\phi\nabla_\nu\phi}$
is a unit vector which is normal to the holographic screen and to
$\xi_\mu$. The Unruh-Verlinde temperature for the metric
(\ref{mat:10}) is simply achieved and reads
\begin{equation}
\label{mat:38}T=\frac{1}{4\pi}\frac{A'}{\sqrt{AB}}.
\end{equation}
On the SSS screen, $N$ bits of information are stored and the
holographic information about the source material is encoded as
$dN=d{\cal{A}}$, where ${\cal{A}}$ is the area of the screen.
According to the equipartition law of energy, the energy $E$ is
distributed on a closed screen of the constant redshift $\phi$. For
further details and for example, we display the energy associated
with a source mass $M$ located at the origin of the coordinate.
According to the figure, the spherical holographic screen $\Omega$
with an equilibrium temperature $T$ and the total equipartition
energy $E$ is placed at a distance of $R$ from the source mass. A
test particle with mass $m$ is located near the screen $\Omega$. The
energy is smoothly distributed over the occupied bits, and is
equivalent to the source mass that would emerge in the part of space
surrounded by the screen. The situation is depicted in
Fig.~\ref{fig:1}.

The energy on the holographic screen $\Omega$ for a generic SSS
spacetime becomes
\begin{equation}
\label{mat:39}E=\frac{1}{4\pi}\int_\Omega e^\phi\nabla\phi
d{\cal{A}}=2\pi r^2T.
\end{equation}
This result is in agreement with the Gauss's law. Using
Eq.~(\ref{mat:38}), the energy on the screen then takes the form
\begin{equation}
\label{mat:40}E=\frac{r^2A'}{2\sqrt{AB}},
\end{equation}
which is exactly the same result as in the pseudotensor scenario
that we have derived from the M{\o}ller definition of energy. This
shows the importance of M{\o}ller's prescription for interpreting
the energy distribution in GR \cite{les}.

\section{\label{sec:4}Summary}
In summary, using Einstein, Landau-Lifshitz, Weinberg, Papapetrou
and M{\o}ller energy-momentum complexes, respectively, we have
computed the energy distributions associated with the SSS
gravitational background. In this way the effective gravitational
mass experienced by a test particle situated at any finite radial
distance in the gravitational field is found. On the other hand,
using the emergent view of gravity we have obtained the emergent
gravitational energy on a SSS screen. Our results show that the
emergent gravitational energy obtained from the Verlinde approach is
identical to the gravitational energy derived from the M{\o}ller
approach. It seems that these two approaches possess similar
behaviors. In both approaches, the energy is in fact contained in a
two-sphere of radius $r$ which gives a taste of the effective
gravitational mass that a test particle experiences. However, the
other prescriptions yield the different results. This may lead to
the approvement of Lessner's argument concerning the significance of
M{\o}ller's prescription in the context of GR.\\

\end{document}